\documentclass[aps,prl,reprint,twocolumn,preprintnumbers,floatfix,nofootinbib]{revtex4-1}

\usepackage[T1]{fontenc} 

\usepackage{graphicx}
\usepackage{epstopdf}
\usepackage{mathrsfs}
\usepackage{amssymb}
\usepackage{verbatim}
\usepackage{color}
\usepackage[dvipsnames]{xcolor}
\usepackage{multirow}
\usepackage{amsmath}
\usepackage{subfig}
 \usepackage{slashed} 
\usepackage{float}
\usepackage[normalem]{ulem}
\usepackage{sidecap}
\usepackage{hyperref}
\usepackage{cancel}
\usepackage{enumerate}
\hypersetup{pdfstartview=FitV,colorlinks=true,linkcolor=blue,citecolor=red,filecolor=black,urlcolor=blue}

\newcommand{\beq}{\begin{equation}}
\newcommand{\eeq}{\end{equation}}
\newcommand{\bea}{\begin{eqnarray}}
\newcommand{\eea}{\end{eqnarray}}

\newcommand{\mrm}{\mathrm}

\def\to{\rightarrow}

\usepackage[parfill]{parskip}
\begin{document}\sloppy 

\preprint{CPHT-RR011.032020}
\preprint{UMN--TH--3913/20, FTPI--MINN--20/03}
\preprint{IFT-UAM/CSIC-20-23}

\vspace*{1mm}

\title{A Model of Metastable EeV Dark Matter}
\author{Emilian Dudas$^{a}$}
\email{emilian.dudas@polytechnique.edu}
\author{Lucien Heurtier$^{b}$}
\email{heurtier@email.arizona.edu}
\author{Yann Mambrini$^{c}$}
\email{yann.mambrini@th.u-psud.fr}
\author{Keith A. Olive$^{d}$}
\email{olive@umn.edu}
\author{Mathias Pierre$^{e,f}$}
\email{mathias.pierre@uam.es}
\vspace{0.5cm}

\affiliation{$^a$ Centre de Physique Th{\'e}orique, {\'E}cole Polytechnique, CNRS and IP Paris, 91128 Palaiseau Cedex, France}
\affiliation{${}^b$ 
Department of Physics, University of Arizona, Tucson, AZ   85721}
\affiliation{${}^c $ Universit\'e Paris-Saclay, CNRS/IN2P3, IJCLab, 91405 Orsay, France}
\affiliation{$^d$William I. Fine Theoretical Physics Institute, School of
 Physics and Astronomy, University of Minnesota, Minneapolis, MN 55455,
 USA}
 \affiliation{
$^e$
Instituto de F\'{i}sica Te\'{o}rica (IFT) UAM-CSIC, Campus de Cantoblanco, 28049 Madrid, Spain} 
\affiliation{
$^f$
Departamento de F\'{i}sica Te\'{o}rica, Universidad Aut\'{o}noma de Madrid (UAM), Campus de
Cantoblanco, 28049 Madrid, Spain}

\date{\today}

\begin{abstract} 
We propose a model where a long-lived pseudoscalar EeV particle can be produced with sufficient abundance so as to account for the cold dark matter density, despite having a Planck mass suppressed coupling to the thermal bath.  Connecting this state to a hidden sterile neutrino sector through derivative couplings, induced by higher dimensional operators, allows one to account for light neutrino masses while having a lifetime that can be much larger than the age of the Universe. 
Moreover, the same derivative coupling accounts for the production of dark matter in the very first instant of the reheating. 
Given the sensitivity of the IceCube and ANITA collaborations, we study the possible signatures of such a model in the form of Ultra-High-Energy Cosmic Rays in the neutrino sector, and show that such signals could be detected in the near future.

\vskip 1cm

\end{abstract}

\maketitle

\setcounter{equation}{0}

\section{I. Introduction}

Despite many efforts, the presence of dark matter (DM) in the Universe \cite{planck} has not yet been confirmed by any direct \cite{XENON,LUX,PANDAX} or indirect \cite{FERMI,HESS,MAGIC,ICECUBE,ANITA} detection signal. Recent limits severely constrain typical WIMP scenarios such as the Higgs-portal \cite{hp,Higgsportal}, $Z$-portal \cite{Zportal}, or even the $Z'$-portal \cite{Zpportal}. 
More complex extensions such as the minimal supersymmetric standard model \cite{mssm,Go1983,ehnos} also have a large part of their parameter space excluded \cite{cmssm,mc12,fittino} from LHC searches~\cite{nosusy} . Direct, indirect and accelerator searches place additional constraints on these models (for a review on WIMP searches and models, see \cite{Arcadi:2017kky}). 
As a consequence, it is important to look for alternative scenarios, including those with ultra-weak couplings such as gravitinos \cite{gravitino,ehnos} or  FIMP's \cite{fimp} (see \cite{Bernal:2017kxu} for a review), highly-decoupled dark sectors \cite{HighlyDecoupled}, or the possibility that DM production occurred in the very early stages of reheating after inflation as in SO(10) grand unification \cite{Mambrini:2013iaa,mnoqz}, anomaly free U(1)' models \cite{Bhattacharyya:2018evo}, spin-2 portals \cite{Bernal:2018qlk}, high scale supersymmetry \cite{Benakli:2017whb,grav2,grav3,gravitino2,highsc,Kaneta:2019zgw}
or even moduli portals \cite{Chowdhury:2018tzw}. In all of these models, it has been shown that the effects of non-instantaneous reheating \cite{Giudice:2000ex,Garcia:2017tuj,grav2} and the non-instantaneous thermalization of reheating products \cite{Garcia:2018wtq} on the production of DM particles are non-negligible.

On the other hand, the absolute stability of DM is usually justified by imposing a symmetry. Discrete symmetries are the most popular (R-parity in supersymmetry \cite{fayet}, a $Z_2$ symmetry in SO(10) unification \cite{DeMontigny:1993gy,mnoqz}, a $Z_2$ symmetry in Higgs \cite{hp, Higgsportal} or Z-portal \cite{Zportal} models) and can arise from broken gauge symmetries which are exact at the Planck scale. This is not the case for continuous global symmetries which are generically violated at the Planck scale \cite{Hawking:1974sw,hpp,Kallosh:1995hi,Banks:2006mm}.
In this case, the decay of DM is rendered possible through Planck-suppressed operators, as argued in \cite{ProfumoPlanck}.

Due to its very specific signature (monochromatic final states for a 2-body decay), a metastable candidate is regularly evoked when specific detection signals are claimed. For example, a positron excess \cite{Profumo:2019pob}, photon lines \cite{Mambrini:2015nza}, high energy neutrinos in IceCube \cite{Dudas:2014bca,neutrinodudas} or ultra-high energy neutrinos in ANITA \cite{Heurtier:2019git,ANITAobs}. However, in each case, the interpretation of a signal as a dark matter detection has to deal with a severe issue: justifying a long lifetime (and thus extremely tiny couplings) while at the same time finding a production mechanism able to produce the dark matter in a sufficiently large amounts to account for the PLANCK determined density of dark matter 
\cite{planck} (implying a coupling which is not so tiny).

At first sight, it would seem that Planck-suppressed couplings of DM particles to Standard Model (SM) states could be sufficient for explaining why dark matter may be long lived on cosmological time scales. For example, one may naively expect the decay width of dark matter to be of order $\Gamma \simeq \frac{m^3}{M_{P}^2}$ where $m$ denotes the DM mass\footnote{We will use the reduced Planck mass throughout the paper, $M_P = \frac{1}{\sqrt{8 \pi G_N}} = 2.4\times 10^{18}$ GeV.}. However, given the 
current limit from indirect gamma \cite{FERMIlimit} positron \cite{Positronlimit} or neutrino \cite{Icecubelimit} detection ($\tau= \Gamma^{-1} \gtrsim 10^{29}$ seconds) one would require  $m \lesssim 10$ keV which reaches the limit from Lyman-$\alpha$ or structure formation constraints~\cite{Palanque-Delabrouille:2019iyz}. It is, moreover, not an easy task to produce 
the requisite abundance of
DM particles with such feeble couplings. Even the FIMP scenario necessitates couplings of the order of $10^{-11}$ \cite{fimp}, i.e., much larger than $\frac{m^2}{M_P^2}$.

A potentially more natural way to couple 
DM to the Standard Model (SM) bath with Planck suppressed couplings is through the neutrino sector, for which there are already strong mass constraints $\sum m_\nu \lesssim 0.15$ eV~\cite{sumnu}. 
Indeed, several constructions invoke a new massive scalar~\cite{Majoronreview} to justify the neutrino mass through a dynamical process similar to the Higgs mechanism applied in the right-handed neutrino sector.

In this work, we show that by combining the violation of global continuous symmetries at the Planck scale, while coupling DM to the neutrino sector, one can generate a large DM lifetime 
\beq
\tau \propto \left(\frac{M_P}{m_\nu} \right)^2 \frac{1}{m} \simeq 10^{33}~\mathrm{s} \left( \frac{0.05~ {\rm eV}}{m_\nu}\right)^2 \left(\frac{1~{\rm GeV}}{m}\right),
\eeq
in compliance with the actual experimental constraints for $\sim$ 1 PeV dark matter masses\footnote{Note that the lifetime of the Universe is $\sim 4 \times 10^{17}$ seconds, corresponding to $\sim 6.5 \times 10^{41}~\mathrm{GeV^{-1}}$ whereas limits from indirect detection reach $\tau \gtrsim 10^{29}$ seconds ($\sim 10^{53}~\mathrm{GeV^{-1}}$).}.     
The paper is organized as follows. In section II we present our model and we compute the DM lifetime and relic abundance in section III. We propose experimental signatures in section IV.  In section V, we
propose a top-down model which incorporates all of the needed components for our EeV DM candidate and its coupling to the observational sector. Our conclusions are given in section VI. Appendix A contains additional details on the computation of the decay rates and Appendix B gives more detail on the UV microscopic model containing
additional particles and interactions which generate, in the IR, the effective model with appropriate mass scales and couplings that we analyze in the bulk of the paper.

\section{II. The model}

\subsection{Motivations}

We begin with some motivation for the general and more detailed models we present below. The models we are proposing rely on a derivative coupling of a DM candidate, $a$,  to matter. 
Indeed, axionic couplings of the type $\frac{\alpha}{M_P}\partial_\mu a$ appears in several ultraviolet constructions. For instance, in models with  string or higher-dimensional inspired moduli fields $T=t+ia$ (see \cite{Chowdhury:2018tzw} for a more detailed study), they can couple to a sterile sector through the kinetic term as 
\beq
{\cal L } \supset \frac{i}{2}[\bar \nu_s \gamma^\mu {\cal Z}_s  \partial_\mu \nu_s - (\partial_\mu \bar \nu_s) \gamma^\mu 
{\cal Z}^*_s \nu_s]
\eeq
with ${\cal Z}_s = 1 + \frac{\beta_s}{M_P}t + i \frac{\alpha_s}{M_P} \gamma_5 a$ and $\alpha_s,\beta_s $ real for simplicity. After an integration by parts, the Lagrangian will contain terms  
\beq
{\cal L} \supset \frac{\alpha_s}{2 M_P} (\partial_\mu a)\bar \nu_s \gamma^\mu \gamma^5 \nu_s \ , 
\label{kincoup}
\eeq
which are of the form we consider below.

We can also find such couplings in the Majoron model. Consider a Lagrangian of the type 
\beq
{\cal L}_\phi = \phi \nu_s \nu_s + {\rm h.c.},
\eeq
written using a two-component notation and where $\phi=\chi e^{\frac{ia}{M_P}}$ is the Majoron. After a redefinition of phases, 
$\nu_s \rightarrow e^{- \frac{ia}{2M_P}} \nu_s$,  the kinetic term,  $-i \bar \nu_s {\bar \sigma}^\mu \partial_\mu \nu_s$, produces a coupling of the type given in Eq.~(\ref{kincoup})~\cite{neutrinobabu,neutrinodudas}.

Even in string constructions, where we can define the moduli superfield in term of the Grassmannian variables $\theta$ and $\bar \theta$
by $T + \bar T = 2 t + 2 \theta \sigma^\mu \bar \theta \partial_\mu a$, we can show that a term $\frac{1}{\langle t^2 \rangle M_P} \partial_\mu a \bar \nu_s \sigma^\mu \nu_s$ appears once expanding the K\"ahler metric as function of matter fields. In this case, $\alpha_s$ can be identified as $\frac{1}{\langle t^2 \rangle} \simeq 10^{-2}-10^{-3}$ in KKLT-like models \cite{Linde:2011ja}. As one can see, several ultraviolet constructions contains couplings of the type (\ref{kincoup}) which we use below. 

\subsection{The Lagrangian}

Our goal in this section is to build a minimal model of metastable EeV dark matter. By minimal, we mean that
we introduce the fewest number of new fields
beyond those in the Standard Model with neutrino masses. 
We assume that DM is a pseudo-scalar field. As alluded to above, the most economical and natural way to proceed is to couple the pseudo-scalar to a sterile neutrino ($\nu_s$) and/or right-handed neutrino ($\nu_R$) sector\footnote{This coupling can be justified in models with large extra dimensions, where SM is localized on a brane, whereas gravity and SM singlets, in particular sterile neutrinos, propagates into a bulk internal space \cite{neutrinoextra} and couple to an axion localized on a distant brane.} as is the case for the pseudo-scalar part of the Majoron. We consider the following Lagrangian
\beq
{\cal L} ={\cal L}_\Phi+{\cal L}_s + {\cal L}_R  ,
\eeq
with 
\begin{equation}
{\cal L}_\Phi= y_f \Phi \bar f f +\Big(y_\phi \Phi \bar{\nu_s}^c \nu_s
+ \mrm{h.c.}\Big)\, ,
\label{Eq:lagrangianphi}
\end{equation}
\begin{equation}
{\cal L}_s = \frac{\alpha}{M_P} 
\partial_\mu a~ \bar \nu_s \gamma^\mu\gamma^5 \nu_s  
-\Big( y_s \tilde{H} \bar L \nu_s +  \frac{1}{2} m_s \bar \nu^c_s \nu_s + \mrm{h.c.}\Big)\, ,
    \label{Eq:lagrangian}
\end{equation}
\begin{equation}
{\cal L}_R= \frac{\alpha}{M_P}\partial_\mu a ~\bar \nu_R \gamma^\mu \gamma^5 \nu_R
-\Big(y_R \tilde H \bar L \nu_R + \frac{1}{2} M_R \bar \nu^c_R \nu_R+\mrm{h.c.}\Big)\, .
\label{Eq:lagrangianr}
\end{equation}
In (\ref{Eq:lagrangian}), we have included a Yukawa coupling, $y_s$ for the sterile neutrino to a $SU(2)_L$ Standard Model doublet, $L$ and the Higgs doublet, $H$, giving rise to a Dirac mass term.
We also include a Majorana mass term for $\nu_s$. In (\ref{Eq:lagrangianr}),
in addition to coupling the pseudo-scalar to $\nu_R$, we add the standard Dirac and Majorana mass terms needed for the see-saw mechanism \cite{seesaw}. 

As it will be important later when we discuss the production of dark matter during reheating, we also introduce an inflaton, $\Phi$, and couple it to both
the sterile sector and the SM, where $f$ corresponds to a SM fermion. 
Finally, $\alpha$ is a coupling $\lesssim 1$ that
represents the physics behind the Planck suppressed terms.

The Lagrangian terms in Eqs. (\ref{Eq:lagrangian}) and (\ref{Eq:lagrangianr}) lead to the following neutrino mass matrix:
\begin{equation}\label{eq:MassMatrix}
\frac{1}{2}
\left(
\begin{matrix}
{\bar \nu}_{L} & {\bar \nu}_{s}^c & {\bar \nu}_{R}^c
\end{matrix}
\right) 
\begin{pmatrix}
 0 & m_D^s & m_D^R \\
m_D^s & m_s &
 0 \\
m_D^R & 0 & M_R
\end{pmatrix}
 \left(
\begin{matrix}
\nu_{L} \\ \nu_{s}^c \\ \nu_{R}^c
\end{matrix}
\right)
+ {\rm h.c.}~,
\end{equation}
where $m_D^s = y_s v_h/\sqrt{2}$, $m_D^R = y_R v_h/\sqrt{2}$ and $v_h = 246$ GeV is the SM Higgs vacuum expectation value. 
We assumed flavor-diagonal couplings in the SM neutrino sector for simplicity and suppressed flavor indices.
We also assume the following mass hierarchy (that will be justified in section V)
\beq
m_s < m_D^R \ll M_R \ .
\label{hier}
\eeq
After diagonalization, we can define the 3 mass eigenstates $\nu_1, \nu_2, \nu_3$ by
\bea
&&
\nu_1= \cos\theta ~(\nu_s + \nu^c_s) + \sin \theta ~(\nu_L + \nu_L^c)
\ , \nonumber
\\
&&
\nu_2 = \cos \theta ~(\nu_L + \nu_L^c) - \sin \theta ~(\nu_s+ \nu_s^c)
\ , \nonumber
\\
&&
\nu_3 \sim \nu_R
\ , \nonumber
\eea
with\footnote{Using the approximation $m_s M_R \ll (m_D^R)^2$}  
\bea
&&
\tan 2 \theta = \frac{2 m_D^s M_R}{(m_D^R)^2 + M_R m_s} \simeq \frac{2 m_D^s}{m_1 + m_2} 
\ , \label{Eq:ys}
\eea
which implies that
\beq
y_s \simeq \sqrt{2} \theta \frac{m_1 +m_2}{v} \simeq \sqrt{2} \theta \frac{m_2}{v} \lesssim 2.9 \times 10^{-13} \theta 
\label{Eq:ys2}
\eeq
and
\bea
&&
m_1 \simeq m_s, ~~m_2 \simeq \frac{(m_D^R)^2}{M_R}, ~~m_3 \simeq M_R \ ,
\eea
where the last inequality in (\ref{Eq:ys2}) assumes a SM-like neutrino mass of $m_2 = 0.05$ eV.

In the $\nu_1, \nu_2$ basis, we can rewrite 
the Lagrangian couplings of $a$ and $\Phi$
to the light neutrino sector as 
\bea
        \mathcal{L} & = & \alpha \dfrac{\partial_\mu a}{M_P} \left( \Bar{\nu}_1 \gamma^\mu \gamma_5 \nu_1 \right. \nonumber \\
        && - \left. \theta ( \Bar{\nu}_2 \gamma^\mu \gamma_5 \nu_1 +  \Bar{\nu_1} \gamma^\mu \gamma_5 \nu_2 ) + \mathcal{O}(\theta^2) \right) \, ,
        \label{an1n2}
\eea 
and 
\bea
        \mathcal{L} & = &  y_{\phi} \Phi \Big( \bar{\nu}_1 \nu_1 - \theta (\Bar{\nu}_1 \nu_2 +\Bar{\nu}_2 \nu_1 ) + \mathcal{O}(\theta^2) \Big) \, .
\eea
As one can see, our framework is similar to a double seesaw mechanism, and the coupling of the dark matter to the standard model will be highly dependent on the mixing angle $\theta$. 
Even if couplings of the form in Eq. (\ref{Eq:lagrangian}) may seem adhoc, they can in fact be justified by high-scale motivated models, an example of which is given in section V.

\section{III. The constraints}

In this section, we consider several necessary constraints on the model. These include 
constraints on the lifetime from indirect detection searches, constraints on the DM abundance - that is we require a viable production mechanism, and cosmological constraints on the sterile sector from contributions to the effective number of neutrino degrees of freedom, $N_{\rm eff}$.

\subsection{Lifetime constraints}

The first constraint we apply to the model is on the lifetime of the dark matter candidate $a$.
To be a viable DM candidate, $a$ should at least live longer than the age of the Universe.  
 However, as was shown in \cite{neutrinodudas}, when dealing with long-lived decays of particles to the neutrino sector, many body final state decays can dominate over two-body decays when a spin flip makes the amplitude proportional to the neutrino mass in the final state. This is reminiscent of three-body annihilation processes generated by internal brehmshtralung which dominate over two-body annihilation processes suppressed for light fermionic final states due to spin-momentum constraints. 

In principle, there are two lifetime limits of importance. First, the DM lifetime (to any final state) must be longer than the age of the Universe ($\tau_a > 4 \times 10^{17}$ s).
Second, the lifetime must exceed $10^{29}$ s
when there is an observable neutrino in the final state \cite{Kachelriess:2018rty,limit2}. 
In our case (see Appendix A for details), the dominant decay channel is indeed the three-body final state decay $\Gamma_{a \rightarrow \nu_1 \nu_2 h/Z}$
and $\Gamma_{a\rightarrow \nu_1 e W}$. All three of these modes have similar amplitudes. Note that we are interested in final states where a SM particle appears, especially an active neutrino, as that gives us the most stringent constraints from experiment\footnote{Note that the dominant 2-body decay has $\nu_1 \nu_1$ in the final state. But for $m_2 \theta > 10^{-5} m_1$, the three body partial width is always larger (see Eq. (\ref{compare}) in Appendix A).}. The $\nu_1 \nu_2 h$ final state is  most important and we obtain (see Appendix A) 
\bea
&&
\Gamma_{a \rightarrow \nu_1 \nu_2 h} = \frac{\alpha^2 \theta^2 m_a^3}{192 \pi^3 v^2 M_P^2}(m_1 + m_2)^2
\ , \label{Eq:lifetime}
\eea
implying that
\beq
\tau_a \gtrsim 5.5 \times 10^{28}\mrm{s}~ \left(\frac{10^{-2}}{\alpha} \right)^2 
\left(\frac{10^{-5}}{\theta} \right)^2
\left(\frac{10^9~ \mrm{GeV}}{m_a} \right)^3
\label{Eq:lifetime2}
\eeq
for $m_1 \ll m_2 \lesssim 0.05$ eV.
Note first the rather amazing result that a pseudo-scalar with mass $10^9$ GeV, has a lifetime which greatly exceeds the age of the Universe. 
This is due primarily to the Planck suppressed coupling and the neutrino mass (squared) in the decay rate. Note also 
that the lifetime of $a$ is determined by the mixing angle $\theta$ which we have normalized
to $10^{-5}$ requiring a relatively small Yukawa coupling of order $10^{-18}$ from Eq. (\ref{Eq:ys2}). The smallness of $y_s$ will be justified in section V. 
In this way, we avoid taking $\alpha$ excessively small\footnote{Indeed, the way we wrote the Planck mass coupling $\frac{\alpha}{M_P}$ imposes $\alpha \lesssim 1$ to avoid large transplanckian BSM scales.}. 

Limits from \cite{Kachelriess:2018rty} gives $\tau_{a\rightarrow \nu_2 \nu_2} \gtrsim 5 \times 10^{28}$ seconds whereas \cite{limit2} obtained 
$\tau_{a \rightarrow b \bar b} \gtrsim 10^{29}$ seconds. To be as conservative as possible, we will consider the upper limit $m_2=0.05$ eV for the neutrino mass and $\tau_a \gtrsim 10^{29}$ seconds throughout our work.

\subsection{Cosmological constraints}
\begin{figure}
\centering 
\includegraphics[width=0.75\linewidth]{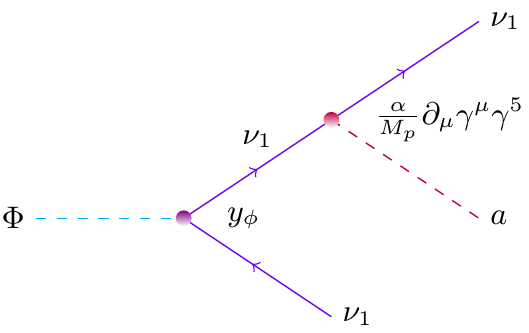}
\caption{\em \small Main inflaton decay channel contributing to the production of dark matter.}
\label{Fig:production1}
\end{figure}

Another important constraint comes from the relic abundance of the dark matter. 
Unless one heavily fine-tunes the coupling of the inflaton to $a$,
the direct production of $a$
through (two-body) inflaton decay
would greatly overproduce the density of $a$ whose annihilation
rate would be extremely small. 
That is, we cannot rely on any kind of thermal freeze-out scenario.
It is however possible to produce $a$ in sufficient quantities through the three-body decay of the inflaton coupled only to SM fermions and the sterile sector as in Eq. (\ref{Eq:lagrangianphi}). This allows for a decay channel $\phi \rightarrow a  \nu_1 \nu_1 $ as shown in Fig.\ref{Fig:production1}.

We assume here some rather generic features of the inflationary sector, and do not need to specify a particular model. We assume a coupling of the inflaton to the SM so that reheating is achieved (we assume instantaneous
reheating and thermalization).
If dominant, the decay rate is given by
\bea
\Gamma_{\Phi \rightarrow {\bar f} f} & = & \frac{y_f^2 N}{8 \pi} m_\Phi \, , \\ \nonumber
\eea
where $N$ is an effective number of final state fermionic degrees of freedom and is similar to the total number of degrees of freedom of the Standard Model. If dominant,
this decay leads to a reheating temperature given by
\beq
T_\text{RH} = \left(\frac{5N}{8\pi^4}\right)^{1/4} y_f \sqrt{M_P m_\Phi}\, .
\label{trh}
\eeq
In general, the relic abundance of dark matter produced in inflaton decay
with a branching ratio $B_R$ can be expressed as \cite{Kaneta:2019zgw}
\beq
\Omega_a h^2 \,\simeq\, 0.1 \! 
\left(\frac{B_R}{9 \times 10^{-16}} \right)
\!\left(\frac{3 \times 10^{13}}{m_\Phi} \right)
\!\left(\frac{T_\text{RH}}{10^{10}} \right)
\!\left( \frac{m_a}{10^{9}}\right)\, ,
\label{Eq:omega1}
\eeq
where all masses are expressed in GeV.
In our specific case, the partial width for producing the DM candidate $a$ is the three body decay width
\beq
\Gamma_{\Phi \rightarrow a \bar \nu_1 \nu_1 } 
= \frac{\alpha^2 y_\phi^2}{24\pi^3} \left(\frac{m_\Phi}{M_P}\right)^2 m_\Phi\, ,
\eeq
and the branching ratio for $\Phi\rightarrow \bar \nu_1 \nu_1 a$ is given by
\bea
B_R & = & \frac{\Gamma_{\Phi\rightarrow \bar \nu_1 \nu_1 a}}{\Gamma_{\Phi \rightarrow \bar f f}} \nonumber \\
& \simeq & \frac{5 \times 10^{-16}}{N} \left(\frac{\alpha}{10^{-2}} \right)^2
\!\left(\frac{y_\phi}{y_f} \right)^2 \!\left(\frac{m_\Phi}{3 \times 10^{13}} \right)^2\!,\nonumber\\
\eea
where we have assumed that the total rate is dominated by the two-body decay to SM fermions, or equivalently that  $N y_f^2\gg y_\phi^2$.

Implementing the expression for the branching ratio into Eq.(\ref{Eq:omega1}) we obtain
\begin{widetext}
\begin{equation}
\Omega_a h^2 \simeq 0.1 \times \frac{125}{N}
\left(\frac{\alpha}{5 \times 10^{-2}} \right)^2
\left(\frac{y_\phi}{y_f} \right)^2
\left(\frac{m_\Phi}{3 \times 10^{13}~\mrm{GeV}} \right)
\left(\frac{T_\text{RH}}{10^{11}~\mrm{GeV}} \right)
\left( \frac{m_a}{10^9~\mrm{GeV}}\right).
\label{Eq:omega2}
\end{equation}
\end{widetext}
We note that the expression (\ref{Eq:omega2}) does not depend on the parameter $\theta$, in contrast to the lifetime of $a$ (\ref{Eq:lifetime}). Indeed, the dominant
decay channel of the inflaton to neutrinos involves only the lighter state, whereas  mixing proportional to $\theta$ is compulsory
for decays with $\nu_2$ in the final state. We also note that to produce the Planck determined  abundance of EeV DM, we need $T_\text{RH} \sim 10^{11}$ GeV when the Yukawa couplings, $y_\Phi$ and $y_f$ are similar\footnote{It is worth mentioning that an alternative possibility would be to produce dark matter only through the graviton-portal but at the price of requiring a very large reheating temperature ($T_\text{RH} \gtrsim 10^{14}$ GeV) as was shown in \cite{Bernal:2018qlk})}. 

\subsection{Constraints on $N_\text{eff}$}

It is also important to consider the contribution of neutrino sector present in our model to the overall expansion rate of the universe. In principle, adding a new light degree of freedom, would increase the effective number of light neutrinos which is strongly constrained by the CMB and BBN.  The current upper limit is \cite{foy2} 
\beq
\Delta N_{\rm eff} < 0.17 \qquad (95\%~ \text{CL}), 
\label{Nefflim}
\eeq
where $\Delta N_{\rm eff} = N_{\rm eff} - 3$.
However, a completely sterile neutrino which would never equilibrate with the SM bath would only contribute a small fraction of a neutrino to $\Delta N_{\rm eff}$ since its energy density gets greatly diluted compared to the energy density of SM neutrinos \cite{sos}.
In cases where a light sterile neutrino (or right-handed $\nu_R$) mixes with the active left-handed neutrinos $\nu_L$, a non-negligible contribution to $N_{\rm eff}$ may result.

Using Eq. (\ref{Nefflim}), we can derive an upper limit on the mixing angle, $\theta$, by noting that interaction rates for $\nu_1$ are the same as those of active neutrinos, $\nu_2$, suppressed by $\theta^2$.
Therefore, $\nu_1$ will decouple 
at a higher temperature, $T_{d1}$, than that
of $\nu_2$, $T_{d2}$.
Indeed we can appoximate $T_{d1} \theta^{2/3} = T_{d2} = 2$ MeV.
As a result, the ratio of the
temperatures of $\nu_1$ and $\nu_2$ at $T_{d2}$ will be given
by 
\beq
\left(\frac{T_1}{T_2} \right)^3 = \frac{43}{4N(T_{d1})}\, ,
\eeq
where $N(T_{d1})$ is the number of degrees of freedom at $T_{d1}$ and the number of degrees of freedom at $T_{d2}$ is 43/4.
Furthermore, the contribution to the number of neutrino degrees of freedom will be 
\beq
\Delta N_{\rm eff} = \left( \frac {T_1}{T_2} \right)^4 = \left( \frac{43}{4N(T_{d1})} \right)^{4/3} .
\eeq
For example, the upper limit in Eq. (\ref{Nefflim}), yields $T_1/T_2 < 0.64$ and $N(T_{d1}) > 162/4$ implying that the decoupling temperature should be greater than $\Lambda_{\rm QCD}$. That is decoupling should take place before the QCD transition in the early universe. Thus we arrive at an upper limit,
\beq
\theta < \left( \frac{T_{d2}}{T_{d1}} \right)^{3/2} = \left( \frac{2 {\rm MeV}}{\Lambda_{\rm QCD}} \right)^{3/2} \lesssim 1.5 \times 10^{-3} \, ,
\eeq
for $\Lambda_{\rm QCD} = 150$ MeV. 

As one can see, in the cosmologically viable region of interest, the value of $y_s$ (and $\theta$) are too weak to be constrained by $N_\text{eff}$ (at the 2$\sigma$ level).
In contrast, the 1$\sigma$ upper limit to $N_{\rm eff}$ is 0.05 \cite{foy2}, and in that case, 
$T_1/T_2 < 0.47$ and $N(T_{d1}) > 407/4$ implying that the decoupling temperature should be as large as $m_t$ (that is greater than all SM masses). In this case, the limit on $\theta$ 
is significantly stronger, $\theta < (2 {\rm MeV}/m_t)^{3/2} \approx 4 \times 10^{-8}$.
Because the number of degrees of freedom varies slowly with temperature above $\Lambda_{\rm QCD}$ the limit on $\theta$ varies quickly with $\Delta N_{\rm eff}$.
At the value of $\theta \approx 1.5 \times 10^{-6}$  corresponding to $\alpha = 0.05$,
we would predict $T_{d1} \approx 15$ GeV $ > m_b$, implying that   $\Delta N_{\rm eff} \approx 0.062$, which may be probed in future CMB missions. In other words, demanding that our model satisfies cosmological constraints therefore favours the region with $\alpha \sim 1$ which is more natural from the model-building point of view.

\subsection{Results}

We note
at this point that the combination of Eqs.~(\ref{Eq:lifetime}) and (\ref{Eq:omega2}) seem to point toward a natural region of the parameter space with $\alpha \simeq 10^{-2}$, $\theta \simeq 10^{-6}$, and $T_\text{RH} \simeq 10^{11}$ GeV which corresponds to $y_\phi \simeq y_{f} \simeq 10^{-5}$ from Eq. (\ref{trh}). In order to explore this region of the parameter space, we performed a scan on the set of parameters $\{y_s, y_R, m_s\}$, while fixing \mbox{$M_R=10^{12}~\mathrm{GeV}$} and requiring that \mbox{$m_2=0.05~\mathrm{eV}$}.

\begin{figure}[ht]
\centering
\includegraphics[width=3.3in]{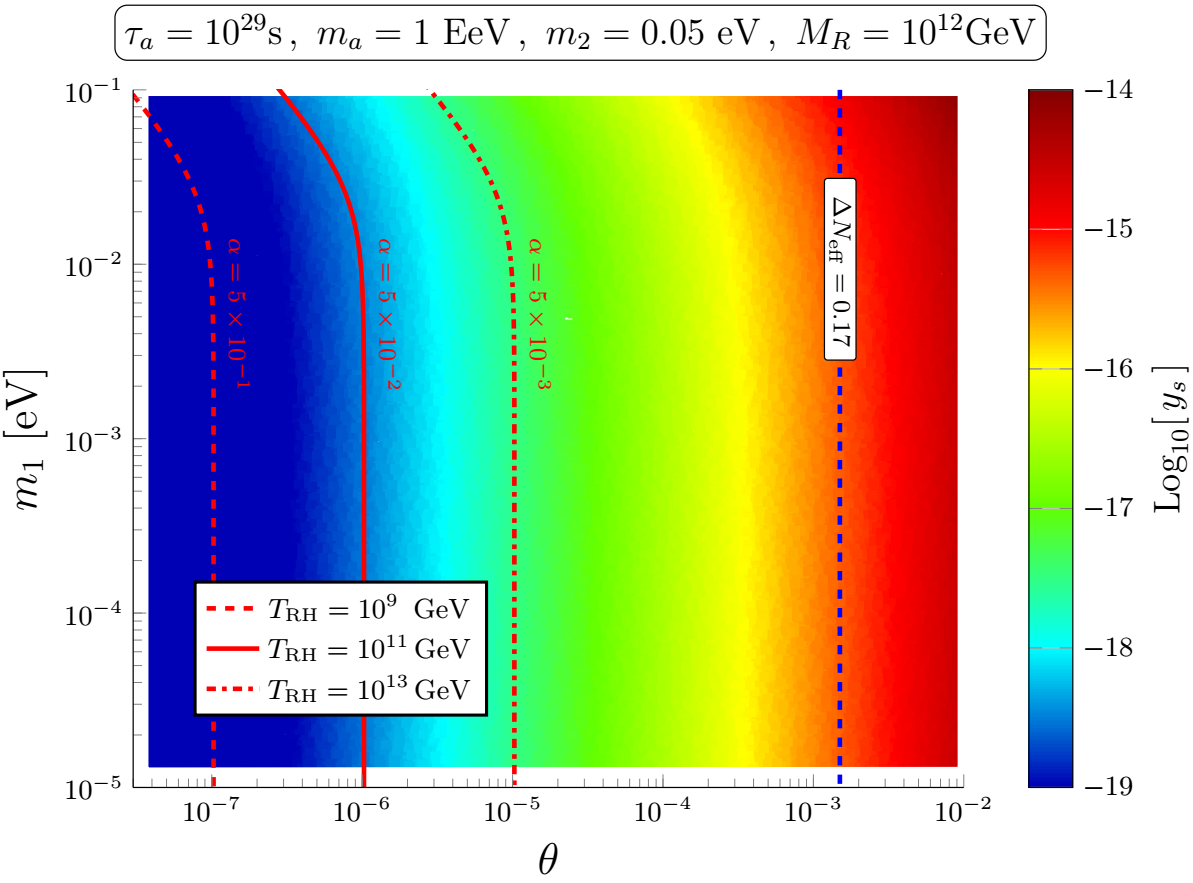}
\caption{\em \small Points in the ($\theta$, $m_1$) parameter space for the case of an EeV dark matter candidate with a cosmological lifetime of $10^{29}$ seconds. The red lines correspond to three different values of $\alpha$ ($5 \times 10^{-1}$, $5 \times 10^{-2}$ and $5 \times 10^{-3}$) and their position in the plane is explained in the text. }
\label{Fig:scan1}
\end{figure}

We show in Fig.~\ref{Fig:scan1} a scan of the plane ($\theta$, $m_1 \approx m_s$) after diagonalization of the mass matrix of  Eq.~\eqref{eq:MassMatrix}. 
For all points considered, we have fixed
the DM mass, $m_a=1~\mathrm{EeV}$, the DM lifetime, $\tau_a =10^{29}~\mathrm{s}$, the active neutrino mass, $m_2 = 0.05$ eV, and the inflaton mass, $m_\Phi = 3 \times 10^{13}$ GeV.
We consider three values of $\alpha$ as labelled. 
In the simple case where the inflaton couples equally to the sterile neutrino $\nu_1$ and SM fermions ($y_\phi = y_f$),
from Eq.~(\ref{Eq:omega2}) we can satisfy  $\Omega_a h^2\simeq0.12$ for different values of $\alpha$ by compensating with a different value the reheating temperature. For the values of $\alpha$ chosen,  we require $T_\text{RH}=10^{13}$, $10^{11}$ and $10^{9}$ GeV
as indicated on the figure. 
The position of the lines of constant $\alpha$ is determined by setting 
the lifetime to the experimental limit of \mbox{$\tau_a =10^{29}~\mathrm{s}$} which can be read from Eq.~\ref{Eq:lifetime} (with $\tau_a \approx \Gamma_{a \rightarrow \nu_1 \nu_2 h}^{-1}$). Note that for $m_1\lesssim m_2$, the lifetime can be approximated by Eq.~(\ref{Eq:lifetime2}) which is independent of $m_1$ which explains why the lines are mostly vertical in the depicted plane. Moreover, since the lifetime is proportional to $(\alpha \theta)^2$, the choice of $\alpha$ determines $\theta$ for constant $\tau_a$.  For each point of the scan, the value of the corresponding Yukawa coupling $y_s$ is indicated by the colored bar.

As was anticipated in the previous subsection, in the region 
of interest the correction to $\Delta N_{\rm eff}$
is quite small as compared to the 95\% CL upper limit limit $\Delta N_{\rm eff}<0.17$, which corresponds to the dashed blue contour in Fig.~\ref{Fig:scan1}.

We also see in the figure, that the coupling $y_s$ should be quite small ($\mathcal{O}(10^{-18})$ in the region of interest). We will justify this small coupling in section V. Note that
the reason the contribution of $\nu_s$ to $N_{\rm eff}$ is small, is precisely 
because the coupling, $y_s$
(and mixing angle) is small.

\section{IV. Signatures}

 \subsection{IceCube signals}

One clear signature of the model discussed above would be a monochromatic neutrino signal that could be observed by the IceCube, or the ANITA collaborations. In Ref.~\cite{Heurtier:2019git}, the case of a scalar DM particle decaying into light right-handed neutrinos was studied and it was shown that the decay of an EeV DM particle followed by the scattering of the RH neutrino within the Earth's crust could lead to visible signals both for ANITA and IceCube for a mixing angle and DM lifetime of order $\tau_{a}/\theta^2\lesssim 10^{27}~\mathrm{s}$. As we have seen, the region of the parameter space that is favored in our model lies towards smaller values of the mixing angle $\theta\lesssim 10^{-5}$ and $\tau_a\gtrsim 10^{29}~\mathrm{s}$, leading to a ratio $\tau_a/\theta^2\gtrsim 10^{39}~\mathrm{s}$. This would indicate that our model cannot be detected in searches for anomalous upward-propagating cosmic rays.

In contrast, searches for downward-propagating ultra-high-energy (UHE) cosmic rays are better suited for signatures of the model discussed here. The IceCube collaboration has reported limits on the decay of dark-matter particles with masses reaching up to a few hundred PeV to active neutrinos. Furthermore, it was shown in Refs.~\cite{Kachelriess:2018rty, Berghaus:2018zso} that the creation of electroweak showers from the decay of a heavy DM state into neutrinos at very high energy might be constrained at lower energy since the secondary products of such a shower might be visible in the form of a diffuse flux of neutrinos or photons at low energy. These studies led us in the previous sections to impose that the DM lifetime is larger than $\tau_a\gtrsim 10^{29}~\mathrm{s}$. In this section we determine the region of parameter space that might be probed experimentally by IceCube, either in the form of direct scattering of UHE neutrinos within the detector, or from secondary electroweak showers which would arrive on Earth at lower energies.

\subsection{Neutrino Scattering in the IceCube Detector}
Let us estimate the number of events which could be detected by IceCube under the form of a monochromatic neutrino signal at ultra-high energies. For that purpose, we suppose that the dark-matter particles follow a  Navarro-Frenk-White (NFW) profile~\cite{Navarro:1995iw} :
\begin{equation}
    \rho_\textrm{DM}(r)\propto \dfrac{1}{\left(\dfrac{r}{r_s} \right) \left[ 1 + \left(\dfrac{r}{r_s} \right)^2 \right]}~,
\end{equation}
where $r_s=24~\text{kpc}$ and the dark-matter density distribution is normalized to equal  $\rho_\odot=0.3~\text{GeV cm}^{-3}$ in the vicinity of the solar system \cite{Catena:2009mf}. 
Following Ref.~\cite{Heurtier:2019git}, the dark-matter flux, averaged over solid angle, is
\begin{equation}
\langle \Phi \rangle \simeq 1.6 \times 10^{-16}~\text{cm}^{-2}~\text{s}^{-1} \left( \dfrac{10^{29}~\text{s}}{\tau_\textrm{DM}} \right)\left( \dfrac{ 1~\text{EeV}}{m_\textrm{DM}} \right)\,.
\end{equation}

The number of events predicted for IceCube, assuming a fiducial volume of $\mathcal V_{\rm IC}\approx(1\mathrm{km})^3$ and an exposure time of $\mathcal T_{\rm exp}=3142.5$ days, is given by the relation
\beq
N_{\rm events}=\mathcal V_{\rm IC} \cdot \rho_{\rm ice}\cdot\mathcal N_{\rm A}\cdot \mathcal{T}_{\rm exp}\cdot \sigma_{\nu N}(E_\nu)\,,
\eeq
where the density of the ice is taken to be $\rho_{\rm ice}=0.92~\mathrm{g cm^{-3}}$, $\mathcal N_{\rm A}$ is Avogadro's constant and we estimate the deep-inelastic scattering cross-section $\sigma_{\nu N}$ of neutrinos scattering off nuclei using the results of Ref.~\cite{Connolly:2011vc}
\beq
\log_{10} \left(\sigma_{\nu N}[E]/\mathrm{cm^2}\right)=\sum_{i=0}^3 p_i\log_{10} \left(E/\mathrm{eV}\right)^i\,,
\eeq
with $p_0=-53.5 (-54.1)$, $p_1=2.66(2.65)$, $p_2=-0.129(-0.112)$ and $p_3=0.00182(0.00175)$ for charged current (and neutral current) interactions, respectively. This yields the final result
\beq
N_{\rm events}\approx 0.2\times\left(\frac{10^{29}\mathrm{s}}{\tau_{a}}\right)\left(\frac{1\mathrm{EeV}}{m_a}\right)\left(\frac{\sigma_{\nu N}[m_a/2]}{2.6\times 10^{-33}\mathrm{cm^2}}\right)\,.
\eeq
Therefore, in the region of the parameter space which we have considered, $\tau_a\gtrsim 10^{29}\mathrm{s}$, the number of events that IceCube might see within the detector is expected to be of order one. Therefore, it is reasonable to suppose that increasing the exposure time by a factor of a few could lead to  the detection of such signal in the relatively near future.

\subsection{Secondary Electroweak Shower Detection}
In Ref.~\cite{Kachelriess:2018rty}, limits on the lifetime of a dark-matter particle decaying into active neutrinos have been derived from IceCube data by studying the secondary showers that would be produced by electroweak states at lower energies. We used the limit of Ref.~\cite{Kachelriess:2018rty} on the lifetime $\tau_a$ as a function of the dark-matter mass $m_a$ in order to translate it into a limit on the mixing angle $\theta$ for a fixed set of parameters.

\subsection{Results}

\begin{figure}[ht]
\centering
\includegraphics[width=3.3in]{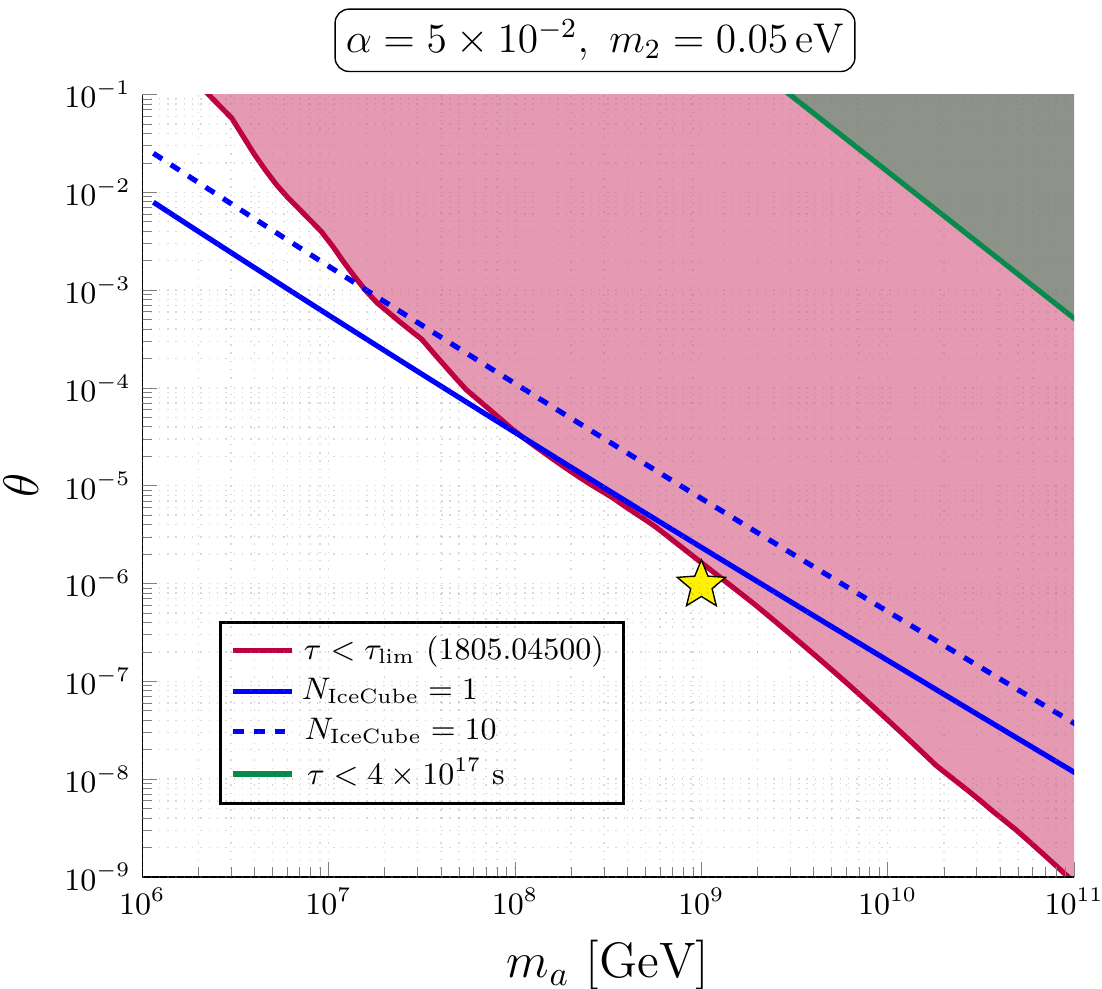}
\caption{\em \small Regions of the $m_a - \theta$ parameter plane that are excluded by astrophysical constraints obtained by neutrino detectors (shaded purple). In the green shaded region, the lifetime of the DM candidate is shorter than the age of the Universe.   Along the solid (dashed) blue line, the number of events expected by Icecube in its exposure time is 1 (10). The yellow star indicates the benchmark point $m_a = 1$ EeV and $\theta \simeq 1.5 \times 10^{-6}$.
}
\label{Fig:icecube}
\end{figure}

\noindent
Our results are summarized in Fig.~\ref{Fig:icecube} where we have fixed the value of $\alpha$ to one of our previous benchmark points, $\alpha=5\times 10^{-2}$, and fixed the active neutrino mass to be $m_2=0.05~\mathrm{eV}\gg m_1$.
The green-shaded area indicates parameter values in the $m_a - \theta$ plane for which the lifetime of dark matter would be shorter than the age of the universe. This occurs only at large values of both $m_a$ and $\theta$ and would be excluded by the lack of events at IceCube. The purple-shaded area excludes the region of the parameter space in which the lifetime of dark matter is shorter than the limit derived in Ref.~\cite{Kachelriess:2018rty}. Finally the blue lines indicate the value of the mixing angle $\theta$ as a function of the DM mass, $m_a$ which would lead to 1 event (solid line) or 10 events (dashed line) in the IceCube detector given the exposure time $\mathcal T_{\rm exp}=3142.5$ days. As one can see, our benchmark point (yellow star) corresponding to $\alpha=5\times 10^{-2}$, $m_2=0.05~\mathrm{eV}\gg m_1$ with $m_a = 1$ EeV and $\tau_a = 10^{29}$ s (corresponding to $\theta \simeq 1.5 \times 10^{-6}$) is flirting with the experimental limits we presented above, suggesting that the prospect for discovery or exclusion of this benchmark is quite high for IceCube, especially for dark-matter masses ranging from the PeV scales to EeV scales.

\noindent
As we have seen, current IceCube data is already on the edge of discovery of EeV dark matter.  In its next phase, starting next year, the IceCube collaboration will be able to probe the EeV scale with much better sensitivity for an observable signal.

\section{V. Towards a microscopic approach}

\begin{figure}[h]
\centering
\includegraphics[width=3.3in]{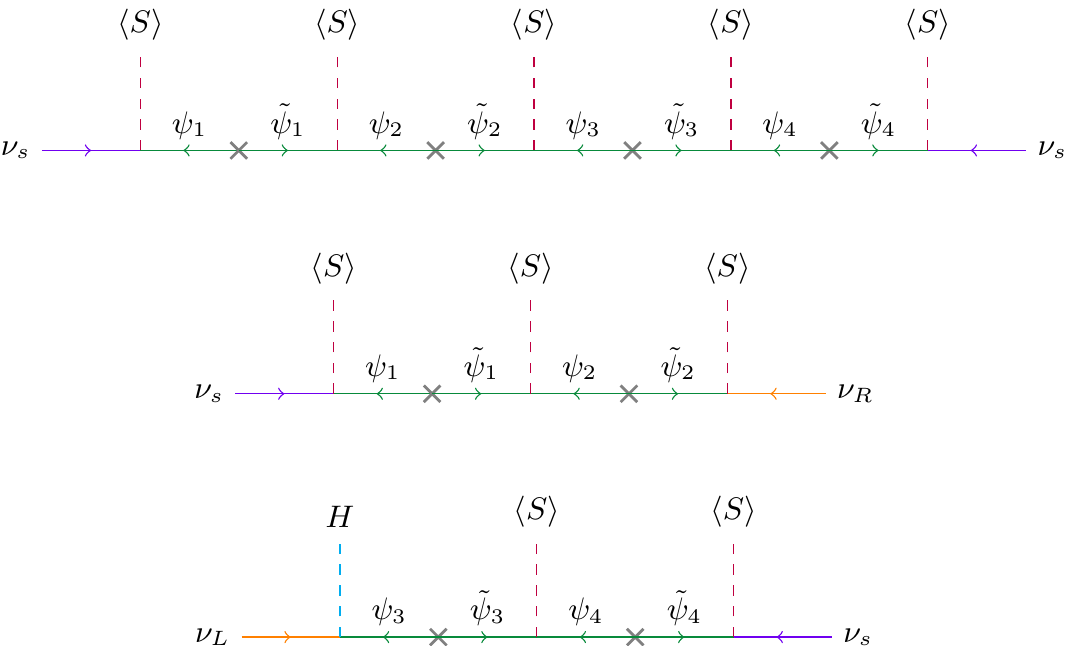}
\caption{\em \small Feynman diagrams involving the heavy fermions $\{\psi_i,\tilde\psi_i\}_{i=1,4}$, the scalar $S$ and the different neutrino species. }
\label{Fig:diagrams}
\end{figure}

In this section we develop a toy microscopic model that could justify our assumed hierarchy given in Eq.~(\ref{hier}).
In fact, such a hierarchy can be generated naturally by the spontaneous breaking of a global $U(1)$ symmetry and the generation of non-renormalizable operators at low-energy. We introduce a set  of heavy Weyl fermion  pairs ${\tilde \psi_i, \psi_i}$ with $i=1,4$ and a complex scalar field $S$ whose charges are given in Table~\ref{tab:charges}. One can integrate out these heavy fermions and obtain effective interactions between the scalar $S$ and the different neutrino species, as can be seen from Fig.~\ref{Fig:diagrams}.
 \begin{center}
\begin{table*}[ht]
    \begin{tabular}{l|c|c|c|c|c|c|c|c|c|c|c|c|c}
      ~ &\quad$a$\quad~& \quad\textbf{$L$~} \quad& \quad\textbf{$S$~} \quad & \quad\textbf{$\nu_s$~~}\quad & \quad\textbf{$\nu_R$~} \quad & \quad $\psi_1$ \quad& \quad $\tilde \psi_1$ \quad& \quad $\psi_2$ \quad& \quad $\tilde\psi_2$ \quad& \quad $\psi_3$ \quad& \quad $\tilde\psi_3$ \quad& \quad $\psi_4$ \quad& \quad $\tilde\psi_4$ \quad\\
      \hline
            \hline
      $U(1)$ &0& -1 & -2 & +5 & +1 & -3&3&-1&1&1&-1&3&-3\\
      \hline
    \end{tabular}
    \caption{\label{tab:charges}\footnotesize Charge assignment of the UV particle content under the new global $U(1)$ symmetry.}
\end{table*}
\end{center}\hfill

Assuming that the fermions $\{\psi_i,\tilde\psi_i\}_{i=1,4}$ have masses of the same order of magnitude $M_i\sim M$ where $M$ is some mass scale close to the Planck scale, after integrating out the heavy fermions, one obtains the effective low energy the Lagrangian
\bea\label{eq:efftheo}
&&
\mathcal L_{U(1)}^{\rm eff} \supset
\frac{\alpha}{M_P} \partial_\mu a ~\bar\nu_s\gamma^\mu\gamma_5\nu_s-
\left(h_S \frac{S^5}{M^4}\bar\nu_s^c\nu_s+h_R S\bar\nu_R^c\nu_R
\right.
\nonumber
\\
&&
\left.
+h_{SR}\frac{S^3}{M^2}\bar\nu_s^c\nu_R+h_{LS}\frac{S^2}{M^2}\tilde H \bar L_L\nu_s+\lambda_L^R \tilde H \bar L_L \nu_R +~\text{h.c.} \right)\nonumber
\\
\label{eff1}
\eea
in four-component notation.  In the above expression,
we have introduced the effective couplings
\bea\label{eq:UVrelations}
h_S=\lambda_4^S\lambda_3^4\lambda_2^3\lambda_1^2\lambda_S^1\,,\quad&& h_{SR}= \lambda_S^1\lambda^2_1\lambda^R_2
\quad
\,,\nonumber\\ h_R = \lambda_R
\quad\text{and} \quad h_{LS}&=&\lambda^3_L \lambda_3^4\lambda_4^S\ , 
\label{eff2}
\eea
where the microscopic couplings $\lambda_i$ are defined in Appendix B. 

We assume that the global $U(1)$ symmetry is broken spontaneously at some high energy scale, and the scalar field acquires a vacuum expectation value $\langle S\rangle\not=0$. After spontaneous symmetry breaking, one obtains the low energy Lagrangian
\begin{eqnarray}
\mathcal L_{\cancel{U(1)}}^{\rm eff} & \supset
\frac{\alpha}{M_P} \partial_\mu a ~ \Bar\nu_s\gamma^\mu\gamma_5\nu_s-\left( \dfrac{1}{2}m_s\bar\nu_s^c\nu_s+\dfrac{1}{2}M_R\bar\nu_R^c\nu_R
\right.
\nonumber
\\
+ &  
\left.
\dfrac{1}{2}m_{SR}\bar\nu_s^c\nu_R+y_s\tilde H \bar L\nu_s+ y_R \tilde H \bar L_L \nu_R + \text{h.c.} \right)  , 
\end{eqnarray}
where we defined
\begin{eqnarray}
m_s&\equiv &2 h_S\frac{\langle S\rangle^5}{M^4}\,,\
M_R\equiv 2 h_R \langle S\rangle\,,\nonumber\\
y_s&\equiv &h_{LS}\frac{\langle S \rangle^2}{M^2}\,, \
y_R\equiv \lambda_L^R\,.
\end{eqnarray}
If we assume all of the couplings $\lambda_i^j \sim 0.1$, with the exception of $\lambda_L^3$ which we take to be $\sim 10^{-4}$, and
a heavy mass scale $M\simeq M_P$ 
and a symmetry breaking scale of
\begin{equation}
\langle S\rangle\simeq 5 \times 10^{12}~\mathrm{GeV}\,,
\end{equation}
we naturally get the desired hierarchy of scales
\bea
m_s\approx 2 \times 10^{-6}~ \mathrm{eV} &\qquad &  M_R\approx 10^{12}~\mathrm{GeV} \nonumber \\  
y_s\approx 4 \times  10^{-18}  & \qquad & y_R\sim 0.1 \ ,
\eea
which approximates the favored parameter space of our model. 
Note that in addition to the model we have previously studied, there is an additional mixing term $h_{SR}\frac{S^3}{M_p^2}\bar\nu_s^c\nu_R$ in the seesaw mass matrix, but we checked that for $h_{SR} \sim 10^{-3}$ this term does not perturb the seesaw mechanism
or our mass hierarchy.   

\begin{figure}[h]
\centering
\includegraphics[width=0.7\linewidth]{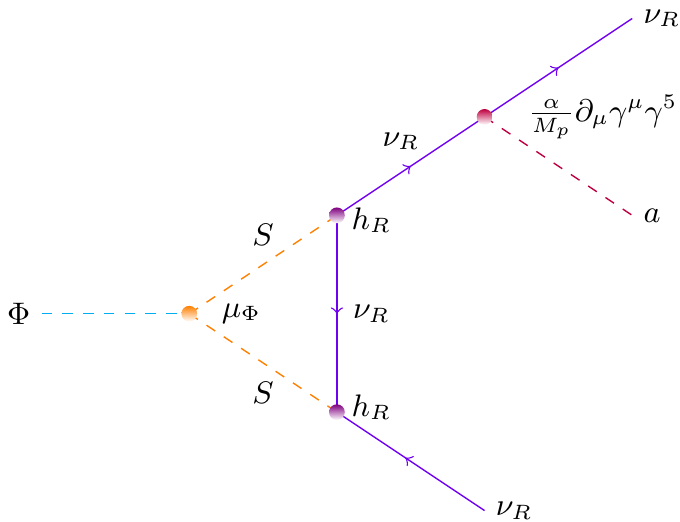}
\caption{\em \small Inflaton decay process, source of the dark matter abundance in the microscopic model.
}
\label{Fig:production2}
\end{figure}

Due to the charge assignment, 
a coupling of the inflaton of the type 
$\Phi \bar \nu_s^c \nu_s$ in Eq.~(\ref{Eq:lagrangianphi}) is not allowed because of the neutrality of $\Phi$ under this $U(1)$. It is then impossible to generate a sufficiently large quantity of dark matter through the decay process shown in  Fig.~\ref{Fig:production1}. However, the production of dark matter via the 3-body decay of the inflaton could be made possible by considering a term like $\mu_\Phi \Phi |S|^2$, with $\mu_\Phi$ being a dimensionful parameter in addition to the term $h_R  S \bar \nu_R^c \nu_R$ with $h_R\sim 0.1$, as depicted in Fig.~(\ref{Fig:production2}). In this case, we expect the effective inflaton decay coupling to be $y_\phi \sim \mu_\Phi h_R^2 f(M_S/M_R)/(16 \pi^2 M_R) \sim 10^{-5}$ for $h_R \sim 0.1$, $\mu_\Phi/M_R \sim 0.1$ and $f(M_S/M_R) \sim 1$ if $M_S \sim M_R$ as we expect, $M_S$ being the mass of the heavy scalar state.

\section{VI. Conclusion}

The most commonly considered mass ranges for dark matter have been either WIMPs with masses between 100 GeV to 1 TeV, or axions with masses much less than an eV. Despite
a vigorous search program neither have yet been discovered. Therefore it is natural to open up the possible mass range for new searches for dark matter.
Indeed there is a lot of effort going into sub-GeV candidates and the prospects for new direct detection experiments. Here we have explored another regime
of dark matter masses of an EeV. 

In this paper, we have shown that we can reconcile the dark matter lifetime, which requires extremely reduced couplings, with a natural production mechanism. The long lifetime is possible when Planck-mass suppressed operators (generated, for example, by the breaking of a global symmetry) are combined with tiny neutrino masses. The induced lifetime respects the strongest indirect detection limits once the dark matter is coupled to the neutrino sector. Moreover, despite the feebleness of the coupling, we showed that inflaton decay into dark matter can be sufficient to produce a relic abundance compatible with PLANCK results. Furthermore, we showed that the next generation of neutrino telescopes will be able to probe such heavy dark matter in the near future.

\vskip.1in
{\bf Acknowledgments:}
\noindent 
The authors want to thank especially Marcos Garcia for very  insightful
discussions. This work was supported in part by the France-US PICS MicroDark and the ANR grant Black-dS-String ANR-16-CE31-0004-01. The work of MP was supported by the Spanish Agencia Estatal de Investigaci\'{o}n through the grants FPA2015-65929-P (MINECO/FEDER, UE),  PGC2018-095161-B-I00, IFT Centro de Excelencia Severo Ochoa SEV-2016-0597, and Red Consolider MultiDark FPA2017-90566-REDC.  The  research activities of L.H. are supported in part by the Department  of  Energy  under  Grant  DE-FG02-13ER41976(de-sc0009913). This work was partially performed at the Aspen Center for Physics, which is supported by NationalScience Foundation grant PHY-1607611. The work of LH has been partially performed during the workshop "Dark Matter as a Portal to New Physics" supported by Asia Pacific Center for Theoretical Physics. KO and MP would like to thank the Lawrence Berkeley National Laboratory for its hospitality during part of the realization of this work. MP would like to thank the Universit\'{e} Libre de Bruxelles for its hospitality during the last stages of the realization of this work. MP and LH would also like to thank the
Paris-Saclay Particle Symposium 2019 with the support of the P2I and SPU research departments and
the P2IO Laboratory of Excellence (program "Investissements d'avenir"
ANR-11-IDEX-0003-01 Paris-Saclay and ANR-10-LABX-0038), as well as the
IPhT. This project has received funding/support from the European Unions Horizon 2020 research and
innovation programme under the Marie Skodowska-Curie grant agreements Elusives ITN No. 674896
and InvisiblesPlus RISE No. 690575. The work of KO was supported in part by the DOE grant DE-SC0011842 at the University of Minnesota.

\section*{Appendix}

\section{A. Decay rates}

In this appendix, we provide some relevant details concerning the computation of the dark matter decay rate. From Eq.~(\ref{an1n2}), we see that up to $\mathcal{O}(\theta^2)$
there are two two-body final state decay channels to consider
\begin{equation}
    \Gamma_{a \rightarrow \nu_1 \nu_1 }=\frac{\alpha ^2 m_a m_1^2 }{\pi  M_P^2} 
    \label{eq:DMwidth}
\end{equation}
\begin{equation}
    \Gamma_{a \rightarrow \nu_1 \nu_2}\simeq \frac{\alpha^2 \theta^2 m_a \left(m_{1}+m_2\right)^2}{2 \pi  M_P^2}
\end{equation}
where we neglected some threshold factors which are negligible in the limit $m_a \gg m_2, m_1$.
The direct decay of $a$ to two SM-like neutrinos is suppressed by $\theta^4$.

There are also three-body final state decays which involve 
a Higgs, $W^\pm$, or $Z$ in the final state.
These couple to neutrinos in the $\nu_1, \nu_2$
basis through
\bea
\mathcal{L}  = &
         - & \dfrac{g}{4c_W} Z^\mu \left( \Bar{\nu_2} \gamma_\mu \gamma_5 \nu_2 \right. \nonumber \\
      & + &  \left. \theta ( \Bar{\nu}_2 \gamma_\mu \gamma_5 \nu_1+  \Bar{\nu}_1 \gamma_\mu \gamma_5 \nu_2 ) + \mathcal{O}(\theta^2) \right) \\
            & - &\dfrac{g}{\sqrt{2}}  \left( \Bar{N_2} \gamma^\mu e_L W_\mu^+ + \Bar{e}_L \gamma^\mu N_2 W_\mu^-  \right. \nonumber \\
           & + & \left. \theta (\Bar{\nu}_1 \gamma^\mu e_L W_\mu^+  + \Bar{e}_L \gamma^\mu \nu_1 W_\mu^- ) + \mathcal{O}(\theta^2) \right)\\
      & - & y_S\dfrac{h}{2\sqrt{2}}\left( \bar{\nu}_2 \nu_1 + \bar{\nu_1} \nu_2 \right. \nonumber \\
      & + & \left. 2\theta( \bar{\nu}_1 \nu_1  - \bar{\nu}_2 \nu_2 ) + \mathcal{O}(\theta^2) \right)  \, ,
\eea
where we used $H=(v_h+h)/\sqrt{2}$ in unitary gauge, where $h$ is the Higgs real scalar field. This leads to 
the three-body decay width with a Higgs in the final which is given by
\begin{align}
    \Gamma_{a\rightarrow \nu_1 \nu_2 h}&=\frac{\alpha^2 m_a^3 y_S^2}{384 \pi^3 M_P^2}=\frac{\alpha^2 m_a^3 \theta^2 }{192 \pi ^3 v_h^2 M_P^2}(m_1+m_2)^2 
\end{align}
where we used the relation between the Yukawa coupling and the mixing angle.
Similarly, 
\begin{equation}
  \Gamma_{a\rightarrow \nu_1 \nu_2 Z}=\frac{\alpha ^2 e^2 \theta ^2 m_a^3 }{768 \pi ^3 M_P^2 c_W^2 m_Z^2 s_W^2}(m_{2}-m_{1})^2 \, ,
\end{equation}
where we assumed $m_{1}\sim m_{2}\ll m_Z \ll m_a$ and $e=gg^\prime/\sqrt{g^2+g^{\prime 2}}$ is the electromagnetic coupling constant. 
Assuming the hierarchy $m_2\sim m_{1}\ll m_e \ll m_W \ll m_a$, gives
\begin{equation}
  \Gamma_{a\rightarrow \nu_1 e_L W}=\frac{\alpha ^2 g^2 \theta ^2 m_a^3 m_{1}^2}{768 \pi^3 M_P^2 m_W^2 } \, ,
\end{equation}
which is of the same order than the partial width $\Gamma_{a\rightarrow \nu_1 \nu_2 Z}$ by using relations between couplings and weak boson masses.

As we have seen, the absence of a helicity flip in the case of the three-body decay compensates the higher power of the Yukawa coupling which arises in the decay width. Thus
the ratio of the 3- to 2-body decay widths is
\bea
        \dfrac{\Gamma_{a\rightarrow \nu_1 \nu_2 h }}{\Gamma_{a\rightarrow \nu_1 \nu_1 }} & = & \frac{m_a^2 \theta^2}{192 \pi ^2 v_h^2} \dfrac{(m_1+m_2)^2}{m_1^2}  \nonumber \\    
        & \gtrsim  & \left( \dfrac{m_2}{m_1}\right)^2 \left( \dfrac{\theta}{10^{-5}}\right)^2 \left( \dfrac{m_{a}}{\text{EeV}}\right)^2 \, ,
        \label{compare}
\eea
and both 2- and 3-body decay modes could be relevant depending on the value of $\theta$
and the ratio of light neutrino masses. However, the 3-body decay always dominates over the 2-body decay when SM particles are in the final state
\begin{equation}
        \dfrac{\Gamma_{a\rightarrow N_1 N_2 h }}{\Gamma_{a\rightarrow N_2 N_1 }}=\frac{m_a^2 }{96 \pi ^2 v_h^2} \gg 1 \, .
\end{equation}

Finally, when going to 4-body and higher decay processes, the major change in the decay width, besides complexifying the phase space volume, is an increase in the powers of the Yukawa coupling
\begin{equation}
\Gamma_{\rm 4-body}\sim \frac{\alpha^2y_S^4}{M_P^2 }m_a^3\quad\Rightarrow\quad \frac{\Gamma_{\rm 4-body}}{\Gamma_{\rm 3-body}}\sim y_S^2\,.
\end{equation}
which naturally leads to smaller widths than the $3-$body decay modes.

\section{B. The microscopic model}

We provide here the microscopic Lagrangian which allows us to derive the $U(1)$ invariant effective theory in Eq.~\eqref{eq:efftheo} of Section V. 
Using a two-component notation, the most general renormalizable and $U(1)$ invariant Lagrangian that one can write involving the fields of the model in Section V is
\begin{eqnarray}\label{eq:LagUV}
\mathcal L_{\rm UV}&\supset &-\left(\lambda_S^1S\nu_s\psi_1+\lambda_1^2S\tilde\psi_1\psi_2+\lambda_2^3 S\tilde\psi_2\psi_3+\lambda_3^4\tilde\psi_3\psi_4\right.\nonumber\\
&+&\left.\lambda_4^SS\tilde\psi_4\nu_s+\lambda^R_2 S\tilde\psi_2\nu_R+\lambda_R S\nu_R\nu_R +\lambda^3_L\tilde H L_L \psi_3\right.\nonumber\\
&+&\left.\lambda_L^R\tilde H L_L\nu_R+\text{h.c.}\right)-\sum_{i=1}^4 M_i\tilde \psi_i\psi_i-V(S)
\nonumber\\
&+& \frac{\alpha}{M_P} \partial_\mu a \Big( \bar{\nu}_s \bar{\sigma}^\mu \nu_s +\bar{\nu}_R \bar{\sigma}^\mu \nu_R \Big)\,.
\end{eqnarray}

Taking a common mass $M_i \equiv M$ for all the super-massive fermions $\{\psi_i,\tilde\psi_i\}_{i=1,4}$, one can integrate them out to obtain the effective operators
\begin{eqnarray}
&&
\mathcal{L}_{\rm eff} \supset 
-\frac{\lambda_4^S\lambda_3^4\lambda_2^3\lambda_1^2\lambda_S^1}{M^4}S^5\nu_s\nu_s 
\nonumber
\\
&&
-\frac{\lambda^R_2\lambda^2_1\lambda_S^1}{M^2}S^3\nu_R\nu_s
-\frac{\lambda^3_L \lambda_3^4\lambda_4^S}{M^2}S^2\tilde H L_L\nu_s
+\text{h.c.}
\end{eqnarray}
One thereafter obtains at low energy the Lagrangian of Eq.~(\ref{eff1}),
written using four-component notation
\begin{equation}
\nu_s\to \begin{pmatrix}
\nu_s\\0
\end{pmatrix}\,,\quad \nu_R\to \begin{pmatrix}
\nu_R\\0
\end{pmatrix}\,,\quad\nu_L\to \begin{pmatrix}
0\\\bar\nu_L
\end{pmatrix}\,,
\end{equation}
and introducing the effective couplings of Eq.~(\ref{eff2}).  

In full generality, certain interaction or mass terms could be added to the Lagrangian of Eq.~\eqref{eq:LagUV} while preserving the symmetries of the model. However, we checked that the presence of such terms do not modify the structure of the effective theory we introduce in Sec. V but simply generate additional contributions to the relations of Eq.~\eqref{eq:UVrelations}.

\vspace{-.5cm}
\bibliographystyle{apsrev4-1}

\end{document}